# Superconductivity in the Heusler Family of Intermetallics

T. Klimczuk[1,2], C. H. Wang[3], K. Gofryk[4], F. Ronning[4], J. Winterlik[5], G. H. Fecher[5], J.-C Griveau[1], E. Colineau[1], C. Felser[5], J. D. Thompson[4], D. J. Safarik[4], and R. J. Cava[6]

[1] European Commission, Joint Research Centre, Institute for Transuranium Elements, Postfach 2340, Karlsruhe, D-76125 Germany

[2] Faculty of Applied Physics and Mathematics, Gdansk University of Technology, Narutowicza 11/12, 80-952 Gdansk, Poland

[3] University of California, Irvine, California 92697, USA

[4] Los Alamos National Laboratory, Los Alamos, NM 87545, USA

[5] Institute for Inorganic and Analytic Chemistry, Johannes Gutenberg-Universität, D-55099 Mainz, Germany

[6] Department of Chemistry, Princeton University, Princeton NJ 08544

## Abstract

Several physical properties of the superconducting Heusler compounds, focusing on two systems (Y, Lu, Sc)$Pd_2Sn$ and $APd_2M$, where A=Hf, Zr and M=Al, In, are summarized and compared. The analysis of the data shows the importance of the electron-phonon coupling for superconductivity in this family. We report the superconducting parameters of $YPd_2Sn$, which has the highest $T_c$ among all known Heusler superconductors.

## Introduction

A hundred years ago, Friedrich Heusler found ferromagnetism in $MnCu_2Al$ [1], a compound that does not contain any ferromagnetic elements. This compound is the prototype of the so called Heusler materials, which crystallize in the cubic $L2_1$ structure and have the general formula $AT_2M$. In the formula, A is generally a transition metal such as Y, Sc, Ti, Hf, Zr, or Nb, but some of the smallest rare earth elements and Mn also form the Heusler phase. T is a transition metal from groups VIIIB or IB of the periodic table, and M is typically an *sp* metal or the metalloids Sb and Bi. More than a hundred ternary intermetallic compounds are known to form in the Heusler structure type, and due to the richness of their physical properties, they are one of the most interesting intermetallic families known. In this class of materials a wide variety of magnetic and electrical transport properties [2,3] including magnetic ordering [4,5] heavy fermion behavior [6,7,8,9,10], shape memory effect [11], half-metallic ferromagnetism [12] and semimetallic [4,5] behavior have been found. Moreover, several Heusler phases have been discovered to have a superconducting ground state (see Refs. [13,14,15,16,17,18,19,20]).

In general, the recipe for finding conventional intermetallic superconductors appears to be simple. One should correctly choose three different metals or metalloids, including a transition metal to ensure a high density of electronic states, to form a new compound. Correctly choosing metals means that one should generally avoid ferromagnetic elements and, according to the Matthias rule [21,22], the ratio of valence electrons/atom should be close to 5 or 7. The Heusler superconductors satisfy this recipe although their superconducting transition temperatures are relatively low. To date, to our knowledge, there are 28 compounds in the Heusler family known to be superconducting.

A full list can be found in Table I. Surprisingly, the coexistence of superconductivity and long range magnetic ordering has also been found in $ErPd_2Sn$ [15] and $YbPd_2Sn$ [20]. In general, despite significant experimental effort, it is still unclear what factor is the most important for superconductivity in Heusler phases. Here, by looking at the common trends of several characteristics such as lattice parameter, Debye temperature, density of states at the Fermi level and electron-phonon coupling in the Heusler phases, we shed more light on this issue. We summarize and compare several physical properties of the superconducting Heusler compounds in the $APd_2M$ family, focusing on two systems (Sc, Lu, Y)$Pd_2Sn$ and $APd_2M$, where A=Hf, Zr and M=Al, In. The analysis of the data shows the importance of the electron-phonon coupling for superconductivity in this family. Moreover, the superconducting parameters of the $T_c$ record holder among the Heusler superconductors, $YPd_2Sn$, are also reported; these support some previous reports [13] and add further information about the phase.

**Experimental**

Polycrystalline samples were prepared by arc melting mixtures of the pure elements in an ultra pure argon atmosphere. Special care was taken to avoid oxygen contamination and therefore a large piece of Zr was used as a getter. The (Sc, Lu, Y)$Pd_2Sn$ and $APd_2In$ (A=Zr, Hf) samples were annealed afterward in evacuated quartz tubes at 750$^o$C and 840$^oC$ respectively. The annealing temperature was held for two weeks before the tubes were quenched in -13°*C* brine. Both resistivity and magnetization tests indicate that while for $APd_2In$ (A=Zr, Hf) the annealing process improves the superconducting properties, for $APd_2Al$ (A=Zr, Hf) the annealing treatment results in a lower, double superconducting transition. Therefore we present the physical properties of unannealed $APd_2Al$ and annealed $APd_2In$.

AC magnetic susceptibility, heat capacity, and *ac* electrical resistivity were measured in a Quantum Design Physical Property Measurement System. For the heat capacity measurements, a standard relaxation calorimetry method was used. For the resistivity measurements we used a standard four-probe technique, with four platinum wires spot-welded to the surface of each, previously polished, sample. DC magnetic measurements of the $YPd_2Sn$ sample were performed using a commercial quantum interference device magnetometer (Quantum Design).

**Results**

The $APd_2M$ (A=Zr, Hf; M=Al, In) and (Y, Lu, Sc)$Pd_2Sn$ samples were characterized before and after annealing by powder X-ray diffraction, carried out on a Scintag XDS 2000 diffractometer with $CuK_\alpha$ radiation ($\lambda$=0.15460 nm). The Heusler compounds crystallize in the cubic $L2_1$ crystal structure (Fm-3m, s.g. 225) and the A atom occupies site 4a (0, 0,0), Pd occupies site 8c ( ¼, ¼, ¼ ) and M occupies site 4b (½, ½, ½ ). All atomic positions are fixed by symmetry. We used the FullProf package [23] to refine the xrd patterns and the cubic lattice parameters *a* obtained are summarized in Table II. These lattice parameters are very close to those reported in the literature. The xrd analysis confirms the good quality of the samples, although the broad diffraction peaks for $HfPd_2Al$ and $ZrPd_2Al$ may suggest either chemical inhomogeneity or difficulty in diffraction sample preparation.

The unit cell size in the (Y, Lu, Sc)$Pd_2Sn$ family depends on the covalent radius of elemental Y, Lu and Sc and is the largest for $YPd_2Sn$ and the smallest for $ScPd_2Sn$. The relative covalent radii of Al and In account for the relative unit cell sizes for $HfPd_2M$ and $ZrPd_2M$ (M=Al, In). The unit cells of the $HfPd_2M$ compounds are smaller than those of the compounds containing Zr ($ZrPd_2M$), likely caused by the slightly smaller covalent radius of the 5*d* metal Hf when compared to the 4*d* metal Zr.

The superconducting transition for $YPd_2Sn$ was first characterized via measurements of dc magnetic susceptibility in the field-cooling and zero-field-cooling mode (10Oe), and are shown in the main panel of Figure 1. In order to estimate the demagnetization factor (*d*), low field magnetization measurement as a function of field $M(H)$ were performed at temperatures 2K, 2.5K, 3K and 3.5K as shown in the inset of Figure 1. At low magnetic fields, the experimental data can be fit with the linear formula $M_{fit} = - aH$. Assuming that the initial linear response to a magnetic field is perfectly diamagnetic ($dM / dH = -1/4\pi$), we obtained a demagnetization factor that is consistent with the sample shape.

In order to estimate lower critical field we followed the procedure used before for $La_3Ni_4P_4O_2$ [24]. The M(H)-$M_{fit}$ data is plotted vs. applied magnetic field (H) in the inset of Figure 2. $H^*$ is the field where M deviates by 2.5% above the fitted line ($M_{fit}$). Taking into account the demagnetization factor, the lower critical field at temperature T, $\mu_0H_{c1}(T)$, can be calculated from the formula $\mu_0H_{c1}(T) = \mu_0H^*(T)/(1-d)$. The main panel of Figure 2 presents $\mu_0H_{c1}$ as a function of temperature. The estimation of $\mu_0H_{c1}(0)$ is possible by fitting experimental data to the formula $\mu_0H_{c1}(T)= \mu_0H_{c1}(0)[1-(T/T_c)^2]$, which is represented by the red solid line. The estimated zero-temperature lower critical field $\mu_0H_{c1}(0)$ = 10 mT, implies a Ginzburg-Landau superconducting penetration depth of approximately $\lambda_{GL}$ = 196 nm. According to our knowledge, these superconducting parameters for $YPd_2Sn$ have not been previously reported.

The superconducting transitions for all four $APd_2M$ (A = Zr, Hf; M = Al, In) samples were characterized by measurements of *ac* susceptibility and electrical resistivity. Figure 3 presents the *ac* susceptibility versus temperature, measured with an applied $\mu_0H_{DC}$ field of 0.5 mT and an applied $\mu_0H_{AC}$ field of 0.3 mT. The left panel (a) presents the superconducting transition for $ZrPd_2Al$ and $HfPd_2Al$. The highest $T_c$ is observed for $HfPd_2Al$, although the double transition suggests inhomogeneity in this sample. Slightly lower $T_c$ is observed for the samples containing In, $HfPd_2In$ and $ZrPd_2In$, which will be discussed later.

The superconducting transition was further examined through temperature dependent measurements of the electrical resistivity ($\rho(T)$). The whole temperature range of $\rho(T)$ for $YPd_2Sn$ is shown in the main panel of Figure 4. The normal state resistivity for $YPd_2Sn$ reveals metallic like character ($d\rho/dT > 0$), although the residual resistivity ratio (RRR) is rather low ~2.5. Such a low RRR is typical for the Heusler compounds, for example the reported value of RRR for $ZrNi_2Ga$ is about 2 [25]. The inset (a) of Figure 4 shows the low temperature resistivity $\rho(T)$ under zero field and applied magnetic fields. A very sharp superconducting transition is observed for 0 and 0.1 T with the superconducting transition width $\Delta T_c$ = 0.2K. Knowing the values of $T_c$ for different magnetic fields [26], we plot the upper critical field values, $\mu_0H_{c2}$ vs. temperature (see the inset (b) of Figure 4). The blue solid line through the data shows the best linear fit with the initial slope $dH_{c2}/dT$ = -0.273 T/K. By using the Werthamer-Helfand-Hohenberg (WHH) [27] expression for a dirty type-II superconductor [28], we estimate the zero-temperature upper critical field $\mu_0H_{c2}(0) = -0.7T_c\ dH_{c2}/dT_c = 0.9$ T for $YPd_2Sn$. This value is comparable with the extracted $H_{c2}(0)$ from Figure 5 in reference [29] and is slightly lower than the 11kOe reported in [13]. With this information, the coherence length can be calculated by using the Ginzburg-Landau formula $\xi_{GL}(0) = (\phi_o/2\pi H_{c2}(0))^{1/2}$, where $\phi_o$=h/2e. The obtained value of $\xi_{GL}(0)$ = 19 nm, and hence the Ginzburg-Landau parameter $\kappa$ = 10, which indicates that $YPd_2Sn$ is a type-II superconductor. Using this parameter, and the relation $H_{c1}H_{c2} = H_c^2 \ln(\kappa)$, we determined the thermodynamic critical field $\mu_0H_c(0)$ = 62 mT.

Figure 5 shows the electrical resistivity in the vicinity of the superconducting transition for $APd_2M$ (A=Hf, Zr, M=Al, In). The highest $T_c$ and a very sharp onset of superconductivity ($\Delta T_c$ < 0.2K) are observed for both $ZrPd_2Al$ and $HfPd_2Al$. Through comparing the $T_c$'s in the group one can infer that Hf and Al promote superconductivity, while Zr and In cause lower $T_c$'s. A double superconducting transition is visible for $ZrPd_2In$. The inset of Figure 5 presents the $HfPd_2Al$ low temperature resistivity ($\rho$(T)) for magnetic fields from 0 to 1.1 T, with a step of 0.1 T. The same procedure as described above for $YPd_2Sn$ was employed in order to calculate the upper critical field ($\mu_0H_{c2}$) for all tested samples. We find the highest upper critical field for $ZrPd_2Al$ and the lowest for $ZrPd_2In$, with the values of 2.82 T and 0.63 T, respectively. The calculated coherence lengths, $\xi_{GL}(0)$, are 11 nm to 23 nm for $ZrPd_2Al$ and $ZrPd_2In$ respectively; these values are comparable to those obtained in the (Y, Lu, Sc)$Pd_2Sn$ family.

The heat capacities measured through the superconducting transitions are shown in the main panel of Figure 6 for $YPd_2Sn$ and Figure 7a for both the Al-containing compounds, $ZrPd_2Al$ and $HfPd_2Al$. The bulk nature of superconductivity is confirmed by sharp, large anomalies at temperatures that are consistent with the $T_c$s determined by the *dc* or *ac* magnetic susceptibility and resistivity measurements. From the temperature dependence of the electronic specific heat ($C_{el}$) below $T_c$ we can extract a value for the superconducting gap by fitting the data to the expected BCS expectation:

$C_{BCS} = t\frac{d}{dT}\int_0^\infty dy\left(\frac{-6\gamma\Delta_0}{k_B\pi}\right)[f\ln f + (1-f)\ln(1-f)]$. Where t = $T / T_c$, f is the Fermi function $f = 1/\left(e^{E/k_BT}+1\right)$, $E = \sqrt{\varepsilon^2 + \Delta^2}$ $y = \varepsilon / \Delta(0)$, and $\Delta(T)/\Delta(0)$ is taken from the tabulated values by Mühlschlegel [30]. The results for $YPd_2Sn$ and $HfPd_2Al$ are shown in inset (a) of figure 6. The gap values are 0.83, 0.59, and 0.51 meV for $YPd_2Sn$, $HfPd_2Al$, and $ZrPd_2Al$ (fit not shown here), respectively. This yields ratios of $\Delta/k_BT_c$ = 2.05, 1.87, and 1.74 respectively, compared with the weak coupling BCS expectation of 1.76, again indicating that $YPd_2Sn$ is the strongest coupling superconductor in the family. This observation is in agreement with what is concluded in ref. [13], where the ratio of $\Delta/k_BT_c$ for $YPd_2Sn$ was calculated to be between 2 and 2.25. The lower than expected value for $ZrPd_2Al$ suggests that the sample is inhomogeneous and does not possess complete superconductivity.

Inset b) of Figure 6 and the panels b) and c) of Figure 7 show the heat capacity measurements, under applied magnetic field, for $YPd_2Sn$ and $APd_2Al$ and $APd_2In$ respectively. The applied magnetic field of $\mu_0 H$ = 3 T was chosen to be above the upper critical field values. The experimental data can be fitted using the formula $C_p = \gamma T + \beta T^3 + \delta T^5$. In this formula the first and two last parameters are the electronic and lattice contributions to the specific heat, respectively. The extracted Sommerfeld coefficients, $\gamma$, are between 6.4 and 10.9 mJ $mol^{-1}$ $K^{-2}$ and are in the range typical of the Heusler materials. Surprisingly in the $APd_2M$ (A=Hf, Zr; M=Al, In) family, the highest $\gamma$ value is obtained for $ZrPd_2In$, the compound with the lowest $T_c$; equally surprising, the lowest $\gamma$ was found for the best superconductor in the series, $HfPd_2Al$, contrary to the naïve BCS expectations.

Using the Sommerfeld coefficient ($\gamma$), and the specific heat jump value at the superconducting transition temperature ($\Delta C$), another important superconducting parameter $\Delta C/\gamma T_c$ can be calculated. Due to low superconducting transition temperature, this calculation was not possible for $ScPd_2Sn$ and $ZrPd_2In$. With one exception, for all other studied compounds $\Delta C/\gamma T_c$ exceeds the BCS predicted 1.426 value, and reaches 1.73 for $YPd_2Sn$ suggesting moderate or strong coupling superconductivity in $YPd_2Sn$. The reason why $\Delta C/\gamma T_c = 1.02$ for $ZrPd_2Al$ compound is unknown, and might be caused by possible inhomogeneity of the superconducting phase as suggested by the broad superconducting transition visible in the *ac* magnetization measurement (see Figure 3 a), and lower than expected $\Delta/k_BT_c$ .

A simple Debye model for the phonon contribution to the specific heat dictates that $\beta$ is related to the Debye temperature through $\Theta_D = \left(\frac{12\pi^4}{5\beta} nR\right)^{1/3}$, where $R = 8.314$ J $mol^{-1}$ $K^{-1}$, and n = 4 is the number of atoms per formula unit. Using the observed values of $\beta$, we find that the Debye temperatures are 182 K and 189 K for $HfPd_2Al$ and $ZrPd_2Al$ respectively. Higher values of the Debye temperature were obtained for $HfPd_2In$ and $ZrPd_2In$, where $\Theta_D$ = 240 K and 235 K, respectively. The similar Debye temperature for the compounds containing either Al ($APd_2Al$) or In ($APd_2In$), are likely due to similar unit cell sizes. The observed trend of $\Theta_D$ deviates from a simple mass relationship – the significantly heavier mass of In should lower $\Theta_D$. A similar surprising behavior is observed in the (Y, Lu, Sc)$Pd_2Sn$ series, in which the Debye temperature for $LuPd_2Sn$ ($\Theta_D$ = 246 K) is higher than that for $YPd_2Sn$ ($\Theta_D$ = 210 K). This suggests the presence of unexpectedly stiff In-Pd and Lu-Pd bonds in $APd_2In$ (A=Zr,Hf) and $LuPd_2Sn$. An even lower value of Debye temperature ($\Theta_D$ = 165 K) for $YPd_2Sn$ was reported in ref. [13].

With these results, assuming $\mu^*$= 0.13 ([31]), the electron-phonon coupling constant ($\lambda_{ep}$) can be calculated from the inverted McMillan's formula [32]:

$$\lambda_{ep} = \frac{1.04 + \mu^* \ln\left(\frac{\theta_D}{1.45T_C}\right)}{(1 - 0.62\mu^*)\ln\left(\frac{\theta_D}{1.45T_C}\right) - 1.04}$$

The observed trend of $\lambda_{ep}$ in the $APd_2M$ (A=Zr, Hf; M=Al, In) series is in agreement with the BCS theory, that is, a stronger electron-phonon coupling causes an increase of $T_c$. Similar behavior is observed for the (Y, Lu, Sc)$Pd_2Sn$ system, which evolves from weak coupling to moderate coupling superconductivity as $\lambda_{ep}$ increases from 0.52 to 0.70. (Using a value of $\mu^*$ of 0.15 causes increase of $\lambda_{ep}$ to 0.75, which is very close to reported 0.79 [13].) Having the Sommerfeld parameter and the electron-phonon coupling, the non-interacting density of states at the Fermi energy can be calculated from: $N(E_F) = \frac{3}{\pi^2 k_B^2 (1 + \lambda_{ep})}\gamma$. The values obtained for all the $APd_2M$ (A=Zr, Hf; M=Al, In) compounds varies from $N(E_F)$ = 2.0 states $eV^{-1}$ per f.u. (formula unit) to $N(E_F)$ = 3.0 states $eV^{-1}$ per f.u, for $HfPd_2Al$ and $ZrPd_2In$ respectively. Samples containing Zr ($ZrPd_2M$) have greater $N(E_F)$, an observation that may be worth investigating by band structure calculations. In the (Y, Lu, Sc)$Pd_2Sn$ family, the lowest ($N(E_F)$ = 1.79 states $eV^{-1}$ per f.u.) and the largest ($N(E_F)$ = 2.23 states $eV^{-1}$ per f.u.) values were obtained for $ScPd_2Sn$ and $YPd_2Sn$, respectively.

## Discussion

With the large number of Heusler superconductors known, it is possible to determine the influence of important materials parameters (i.e. the lattice constant, Debye temperature, Sommerfeld parameter and electron-phonon coupling constant) on the superconducting critical temperature, $T_c$. Using available data, this can be done for 7 superconductors in the $APd_2M$ (A=Zr, Hf; M=Al, In) and (Sc,Lu,Y)$Pd_2Sn$ families.

Figure 8a shows $T_c$ versus lattice constant $a$ for the Heusler superconductors. For (Y, Lu, Sc)$Pd_2Sn$ a larger unit cell causes an increase of $T_c$. This has been discussed previously [33], where it was used to introduce partial atomic disorder in $YPd_2Sn$ in order to increase the lattice parameter, and as a result a higher $T_c$ = 5.5K was observed for $Y_{0.96}Pd_{2.08}Sn_{0.96}$. The same trend, but in the opposite direction, has also been discussed [14], where a negative effect on $T_c$ with applied hydrostatic pressure was reported. The authors of ref. 14 suggest that the depression of $T_c$ in $RPd_2Z$ (R=Sc, Y, Tm, Yb, Lu, and Z = Sn, Pb) is due to a stiffening of the Pd sublattice with increasing pressure. The lattice parameter for the $APd_2M$ (A=Zr, Hf; M=Al, In) system is smaller compared to (Y, Lu, Sc)$Pd_2Sn$. $ZrPd_2In$ has the largest lattice parameter in the first system, comparable to the one for $ScPd_2Sn$, which has the smallest $a$ in the latter system. Interestingly, $T_c$ vs. $a$ in the $APd_2M$ (A=Zr, Hf; M=Al, In) system shows the opposite trend: decreasing the size of the unit cell causes an increase of $T_c$. Although such conflicting trends may reflect the presence of a sharp feature in the electronic density of states that results in an unexpected lattice size dependence of $T_c$, further experimental effort, such as studying the transition temperature under applied pressure for $HfPd_2Al$ and $ZrPd_2Al$ would be of interest, as would further theoretical consideration of this family.

The Debye temperature influences $T_c$ in the same, although unexpected, fashion in the whole series. The BCS theory predicts that $T_c$ should increase with increasing frequency of the lattice vibrations. For the Heusler phases, however, as is shown in Figure 8b, $T_c$ decreases with the Debye temperature.

The next figure (Fig. 8c) presents the superconducting transition temperature versus the density of states at the Fermi energy, $N(E_f)$. For (Y, Lu, Sc)$Pd_2Sn$ subsystem (data represented by close triangles) increasing $N(E_f)$ rapidly increases $T_c$. Again the opposite trend is visible for $APd_2M$ (data represented by close circles). We conclude that $T_c$ changes in a different way depending on the subsystem, similar to what is observed in Fig. 8a.

The electron phonon parameter, $\lambda_{ep}$, is expected to increase $T_c$ as well within the BCS explanation of intermetallic superconductors. This parameter is the one that unifies all the observations in the Heusler family of superconductors. Clear relation $\ln T_c$ vs $-1/\lambda_{ep}$, expected by McMillian formula [32], is shown in Figure 9 for all 7 studied compounds in this family. In the inset of Figure 9, additional data points (open triangles) are shown. In particular, the low $T_c$ (and low $\lambda_{ep}$) points are for the Ni based Heusler compounds for which $\lambda_{ep}$ values have reported [19, 25]. These points fall on the McMillian relation drawn in the main panel of Figure 9. It is worth noting that although the electron phonon coupling has been calculated from the McMillian formula, and therefore $T_c$ depends on $\lambda_{ep}$, there is another variable (Debye temperature) in the formula, that is different for all compounds. Several strong coupling superconductors with much higher $T_c$s, follow the same trend ($Cu_{1.86}Mo_6S_6$ and $Nb_3Sn$).

## Conclusions

A full list of 28 superconductors in the Heusler family, divided into groups with the same number of valence electrons (N), is presented in Table 1. Figure 10 presents the superconducting critical temperatures vs. N per atom. Most of the compounds (19 members) belong to the group with 27 valence electrons per formula unit, which is equal to 6.75 electrons/atom. The record holder is $YPd_2Sn$, with $T_c$ = 4.7 K (although one group reports a $T_c$ with the highest value of 5.5K for non-stoichiometric $Y_{0.96}Pd_{2.08}Sn_{0.96}$ [33].) The blue solid line in Figure 10 shows the trend in the $YAu_{2-x}Pd_xIn$ system, in which Au atoms can be fully replaced by Pd atoms, resulting in a continuous change of the valence electrons from 26 to 28 [18]. Seven different compositions were studied and their $T_c$s are shown as the stars on the Figure. Fifty years ago, Matthias proposed that the superconducting critical temperature of pure elements has a maximum for the ratio of valence electrons / atom slightly below 5 [21]. In a subsequent paper, he proposed the existence of two maxima, close to 5 and 7 valence electrons / atom [22]. Figure 10 suggests that the Heusler superconductors follow this empirical rule, with the most superconductors found at 6.75 electrons/atom, though the fact that superconductors are found for a range of electron counts indicates that electron count is not a hard parameter for determining $T_c$ in this family. Surprisingly, there is only one data point for 6.5 electrons/atom on Figure 6; thus that part of the family is not well characterized. The low $T_c$ of the 6.5 electrons/atom compound suggests that this would not be a fruitful electron count to check for higher $T_c$ Heusler superconductors, but before such a conclusion can be firmly drawn more compounds should be synthesized and tested. If the Heusler phase can be made stable at lower electron counts, then it would be of interest to check those materials for superconductivity to determine whether this family fully follows Matthias' empirical two peak rule for intermetallic superconductors. Further, given the simplicity of the Heusler crystal structure, the large number of superconductors it hosts at different electron counts, and the clarity of the relationship between $T_c$ and $\lambda$ presented in Fig. 9, detailed theoretical modeling of this family of superconductors may be of significant interest.

**Acknowledgement**

The work at Los Alamos National Laboratory was performed under the auspices of the U.S. Department of Energy, Office of Science. The work at Princeton University was supported by the US Department of Energy, grant DE-FG02-98ER45706. The work at the Institute of Transuranium Elements was supported by the grant holder contract. TK acknowledges the European Commission for financial support in the frame of the "Training and Mobility of Researchers" programme.

## Tables

Table I: List of the known Heusler phase superconductors sorted with respect to the number of valence electrons per formula unit. The chemical formula is followed by the superconducting $T_c$ and then by the reference.

| **# 26** | $T_c$ (K) | **# 27** | $T_c$ (K) | **# 28** | $T_c$ (K) | **# 29** | $T_c$ (K) |
|---|---|---|---|---|---|---|---|
| $YPd_2In$ | 0.85 [13]<br>1.04 [18] | $ScPd_2Sn$ | 2.0 [a)]<br>2.05 [34] | $YAu_2In$ | 1.74 [18] | $NbNi_2Sn$ | 2.9 [19]<br>3.4 [18] |
| | | $YPd_2Sn$ | 3.72 [13]<br>4.55 [34]<br>4.7 [a)]<br>5.5 [35] | $ScAu_2Al$ | 4.4 [36] | | |
| | | $LuPd_2Sn$ | 2.8 [a)]<br>3.05 [34] | $ScAu_2In$ | 3 [36] | | |
| | | $TmPd_2Sn$ | 2.82 [34] | $YPd_2Sb$ | 0.85 [13] | | |
| | | $YbPd_2Sn$ | 2.46 [20] | $NbNi_2Al$ | 2.15 [19] | | |
| | | $ErPd_2Sn$ | 1.17 [15] | $NbNi_2Ga$ | 1.54 [19] | | |
| | | $ZrPd_2Al$ | 3.2 [37]<br>3.4 [a)] | | | | |
| | | $ZrPd_2In$ | 2.19 [a)]<br>3.1 [37] | | | | |
| | | $HfPd_2Al$ | 3.66 [a)]<br>3.8 [37] | | | | |
| | | $HfPd_2In$ | 2.4 [37]<br>2.86 [a)] | | | | |
| | | $ZrNi_2Ga$ | 2.9 [25] | | | | |
| | | $ZrNi_2Al$ | 1.38 [18] | | | | |
| | | $HfNi_2Ga$ | 1.12 [18] | | | | |
| | | $HfNi_2Al$ | 0.74 [18] | | | | |
| | | $ScPd_2Pb$ | 2.4 [17] | | | | |
| | | $YPd_2Pb$ | 2.3 [17]<br>4.76 [13] | | | | |
| | | $TmPd_2Pb$ | 2.1 [17] | | | | |
| | | $YbPd_2Pb$ | 2.8 [17] | | | | |
| | | $LuPd_2Pb$ | 2.4 [17] | | | | |

a) – this work

Table II: Characterization of the superconductivity in the (Y, Lu, Sc)$Pd_2Sn$ and $APd_2M$ families of Heusler compounds for A=Zr, Hf; M=Al, In.

| | $YPd_2Sn$ | $LuPd_2Sn$ | $ScPd_2Sn$ | $HfPd_2Al$ | $ZrPd_2Al$ | $HfPd_2In$ | $ZrPd_2In$ |
|---|---|---|---|---|---|---|---|
| $T_c$ (K) | 4.7 | 2.8 | 2.0 | 3.66 | 3.40 | 2.86 | 2.19 |
| a (Å) | 6.7160(8) | 6.6401(3) | 6.5021(8) | 6.3728(7) | 6.3942(9) | 6.5342(4) | 6.5534(5) |
| $\gamma$ (mJ/mol $K^2$) | 9.2(2) | 7.4(1) | 6.6(2) | 7.9(3) | 9.0(1) | 8.5(2) | 10.9(2) |
| $\Theta_D$ (K) | 210(4) | 246(2) | 277(1) | 182(3) | 189(1) | 243(5) | 236(5) |
| $\Delta C/\gamma T_c$ | 1.73 | 1.45 | --- | 1.50 | 1.02 | 1.72 | --- |
| $\lambda_{ep}$ | 0.70 | 0.58 | 0.52 | 0.68 | 0.65 | 0.58 | 0.55 |
| $N (E_F)$ (states/eV/f.u.) | 2.23 | 1.99 | 1.79 | 2.0 | 2.32 | 2.27 | 3.0 |
| $\Delta$ (meV) | 0.83 | --- | --- | 0.59 | 0.51 | --- | --- |
| $\Delta / k_BT_c$ | 2.05 | --- | --- | 1.87 | 1.74 | --- | --- |
| $\mu_0H_{c1}$ (mT) | 10 | --- | --- | 9 | --- | --- | --- |
| $\mu_0H_{c2}$ (T) | 0.90 | 0.45 | 0.26 | 1.81 | 2.82 | 1.00 | 0.63 |
| $\mu_0H_c$ (mT) | 62 | --- | --- | 76 | --- | --- | --- |
| $\xi_{GL}$ (nm) | 19 | 27 | 36 | 13 | 11 | 18 | 23 |
| $\lambda_{GL}$ (nm) | 196 | --- | --- | 225 | --- | --- | --- |
| $\kappa$ | 10 | --- | --- | 17 | --- | --- | --- |

**Captions**

Figure 1

(Color online) Zero-field cooling (ZFC) and field cooling (FC) *dc* susceptibility versus temperature for $YPd_2Sn$. The inset shows field dependent magnetization data M(H) at constant temperatures of 2K, 2.5K, 3K and 3.5K. The black line corresponds to a linear relation (~H) below 60Oe.

Figure 2

(Color online) Temperature dependence of the lower critical field ($\mu_0 H_{c1}$) obtained from magnetic susceptibility. The red line through the data points is the fit as explained in the main text. The inset shows deviation from a fitted linear dependence on H.

Figure 3

(Color online) Temperature dependent ac-susceptibility characterization of the superconducting transitions for a) $APd_2Al$ and b) $APd_2In$ where A=Zr, Hf.

Figure 4

(Color online) Electrical resistivity in a wide temperature range for $YPd_2Sn$. The inset a) shows resistivity measured near the superconducting transition for applied magnetic fields. The inset b) presents the upper critical field ($\mu_0 H_{c2}$) from resistivity as a function of temperature.

Figure 5

(Color online) Electrical resistivity near the superconducting transitions of $APd_2M$, A=Zr, Hf; M=Al, In, under zero field. The inset shows the $HfPd_2Al$ low temperature resistivity measured for applied magnetic fields. Superconductivity is not observed above 1.8K for $\mu_0H > 1.5T$.

Figure 6

(Color online) Zero field specific heat divided by temperature ($C_P/T$) versus temperature for $YPd_2Sn$. The inset a) shows electronic specific heat $C_{el}$ (in a logarithmic scale) vs. $T_c/T$ for $YPd_2Sn$ (green open circles) and $HfPd_2Al$ (blue open squares). The lines represent the BCS fit as explained in the text, which allows us to estimate the superconducting gap value. The inset b) presents $C_p/T$ as a function of temperature, under applied magnetic field, for $YPd_2Sn$. The red line is the $C_p/T = \gamma + \beta T^2 + \delta T^4$ fit at low temperature range.

Figure 7

(Color online) Left panel (a): zero field specific heat divided by temperature ($C_P/T$) versus temperature for $APd_2Al$, where A=Zr, Hf. Right panel: specific heat divided by temperature ($C_P/T$) versus $T^2$, measured under magnetic field $\mu_0H$=3T, for (b) $APd_2Al$ and (c) $APd_2In$ where A=Zr, Hf. The solid line is the $C_p/T = \gamma + \beta T^2 + \delta T^4$ fit in the low temperature range.

Figure 8

(Color online) Superconducting critical temperature, $T_c$, versus lattice parameter (a), Debye temperature (b), and density of states at the Fermi energy (c), for both $ARh_2M$ and $APd_2M$ systems. The dotted and solid lines show the trend for the $APd_2Sn$ (A=Y, Lu, Sc) and $APd_2M$ (A=Zr, Hf, M=Al, In) systems, respectively.

Figure 9

(Color online) Logarithm of superconducting critical temperature, $T_c$, versus inverted electron-phonon coupling parameter, $-1/\lambda_{ep}$, for both (Y, Lu, Sc)$Pd_2Sn$ and $APd_2M$ (A=Zr, Hf, M=Al, In) systems. The dashed line emphasizes the observed trend. In the inset, additional data points for selected intermetallic superconducturs (open triangles) are shown. The low $T_c$ (and low $\lambda_{ep}$) points inside a square are for the Ni based Heusler compounds (ref. [19, 25]).

Figure 10

(Color online) Superconducting critical temperature, $T_c$, versus the number of valence electrons per atom for all known Heusler superconductors. Data points taken from cited references, open symbols from reference [37], stars from reference [18].

**Figures**

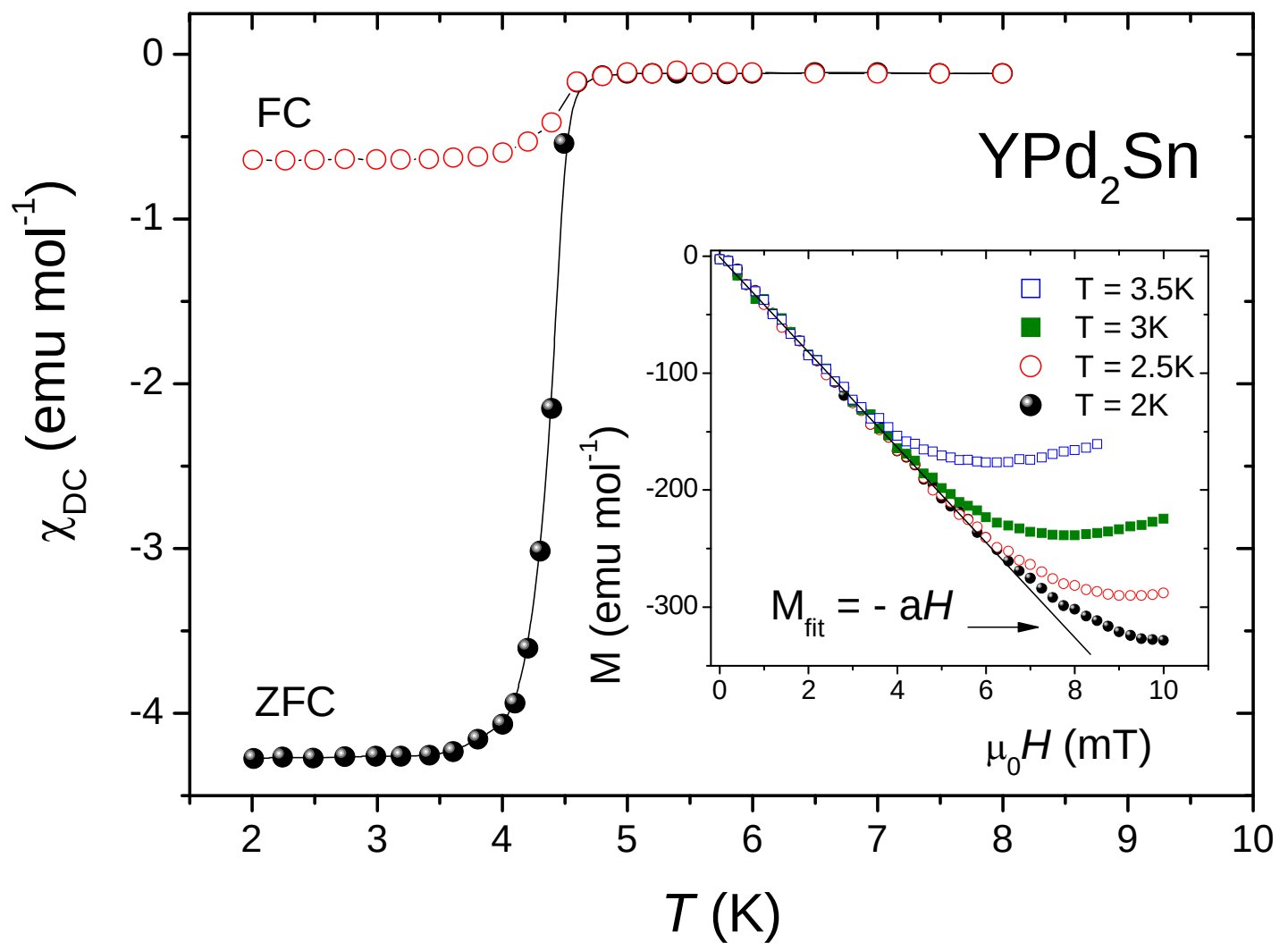

Figure 1.

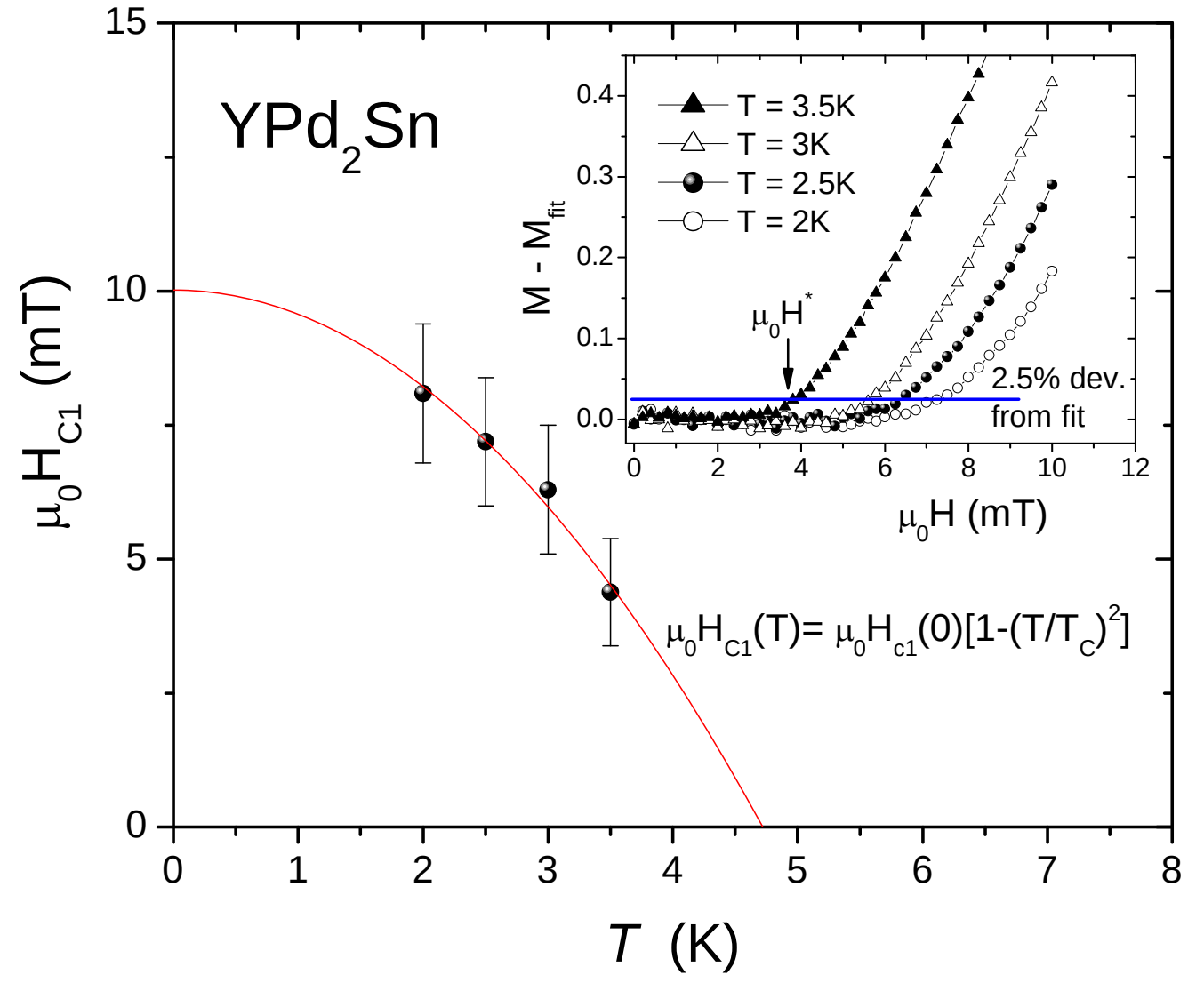

Figure 2.

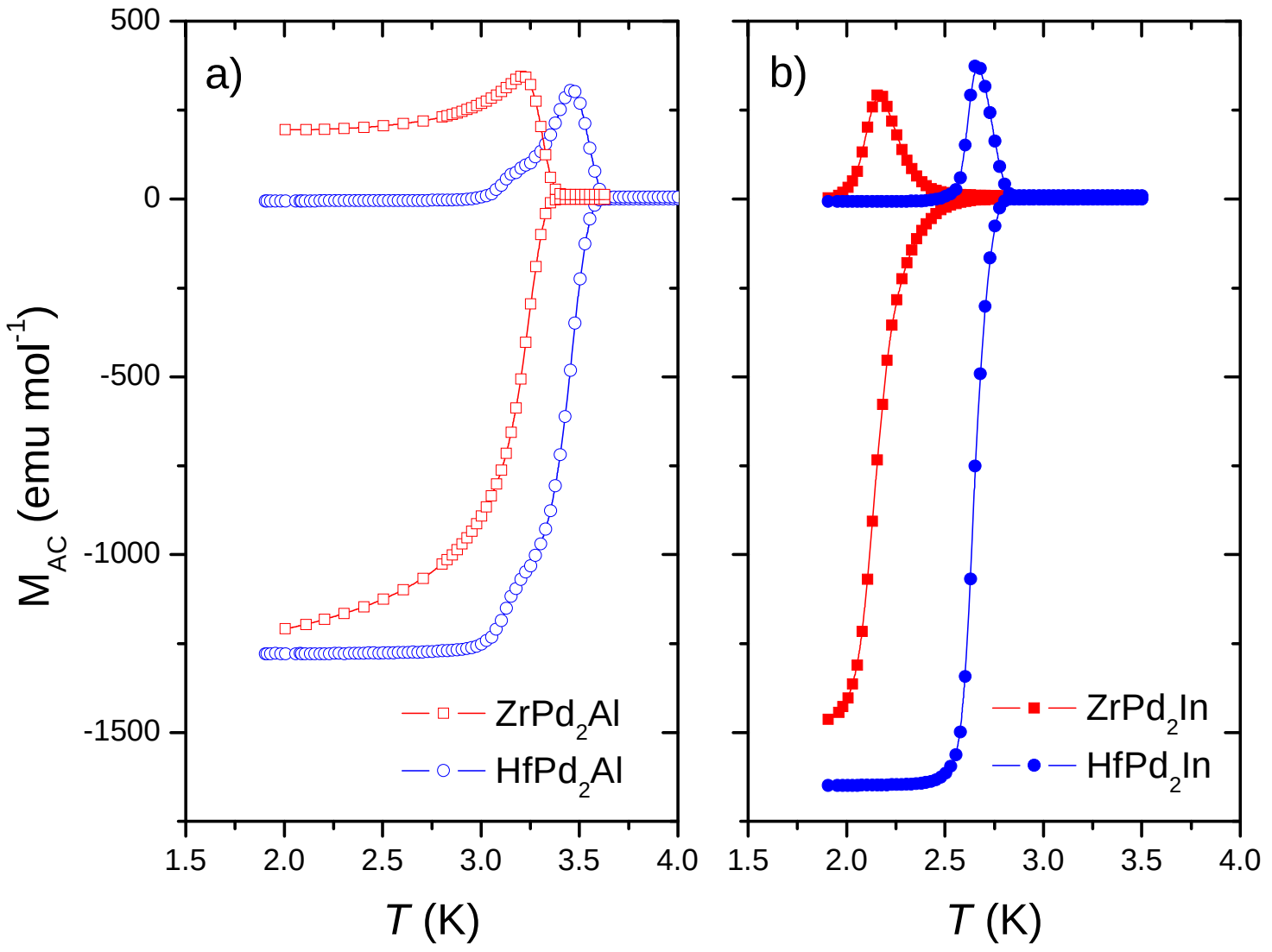

Figure 3.

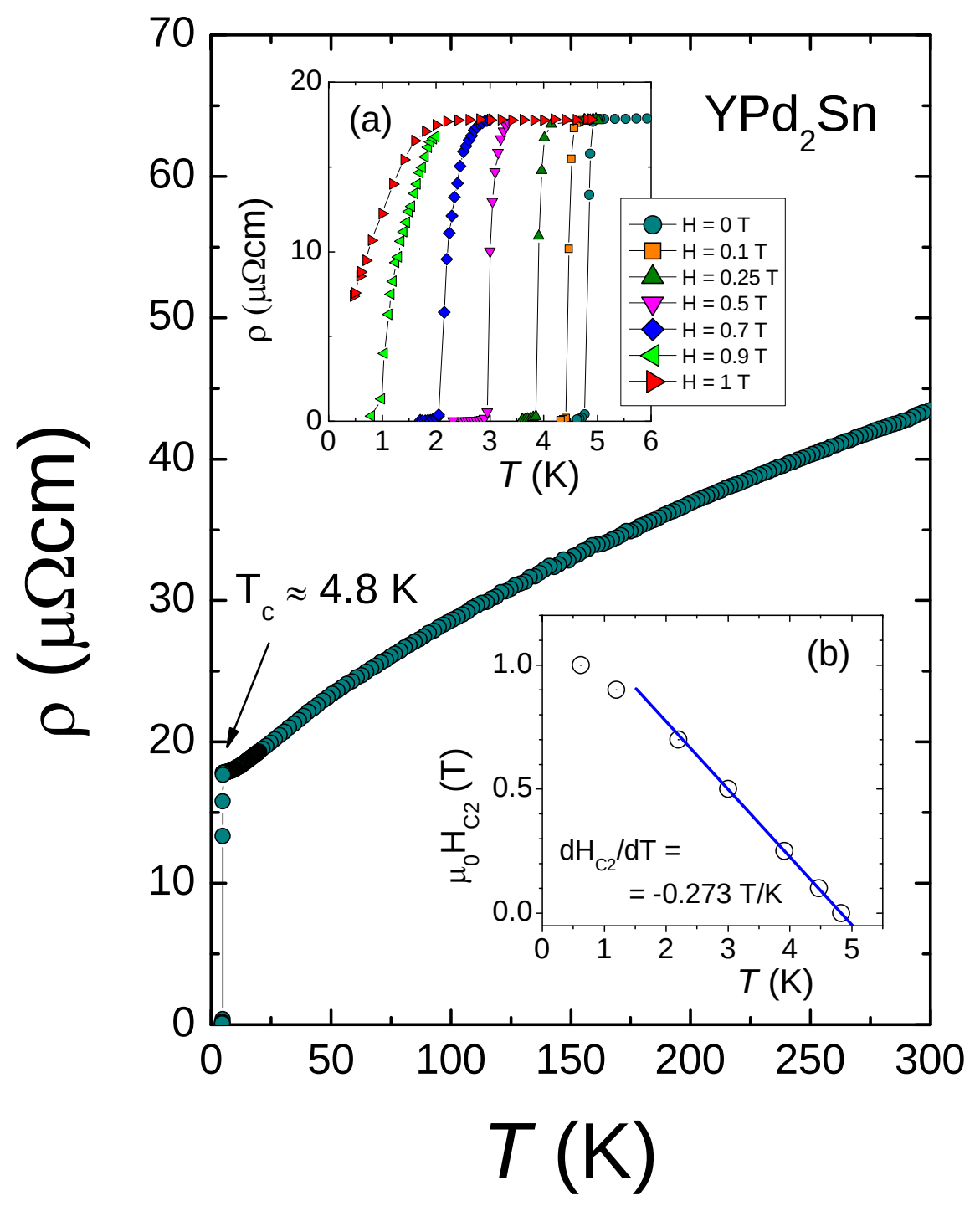

Figure 4.

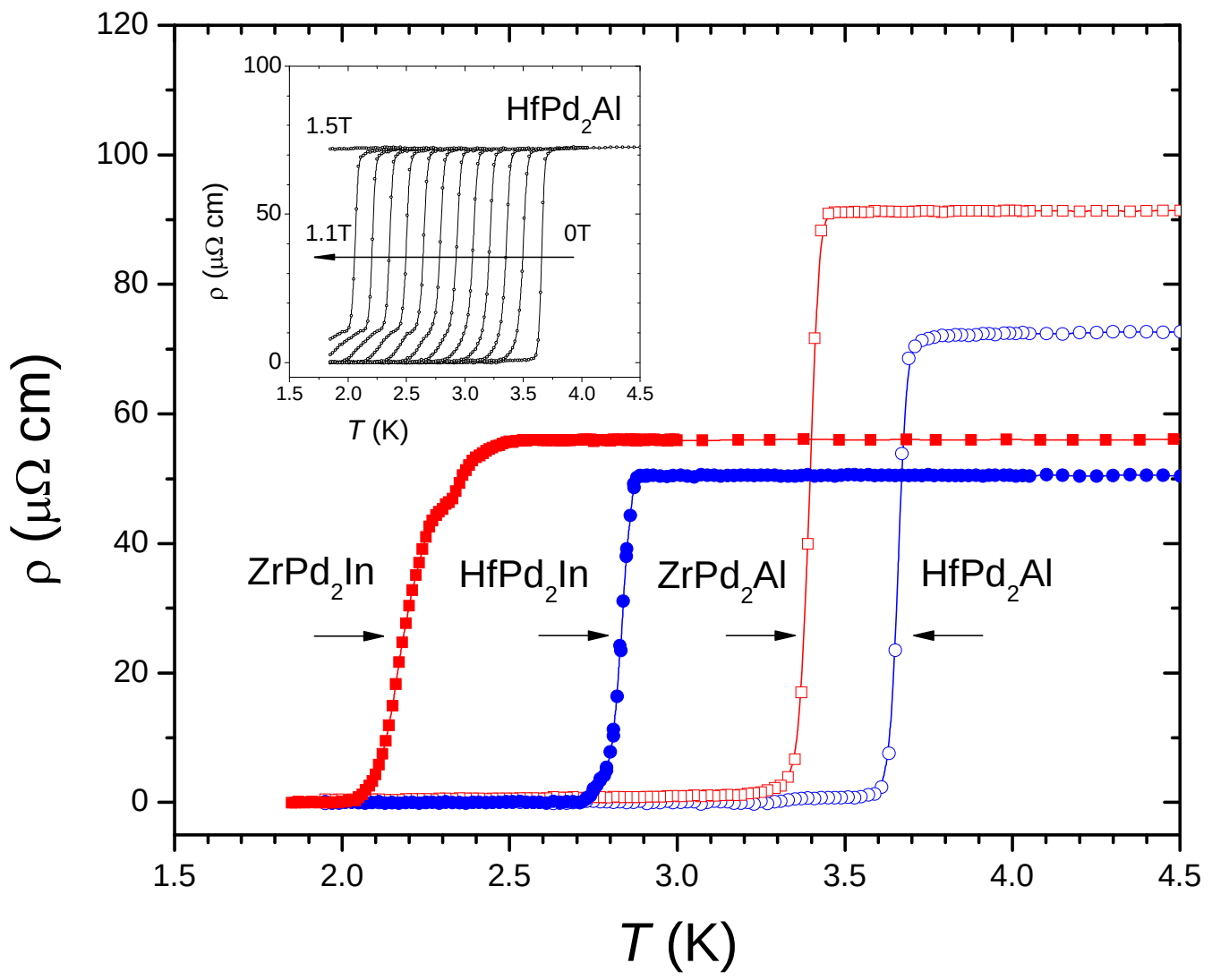

Figure 5.

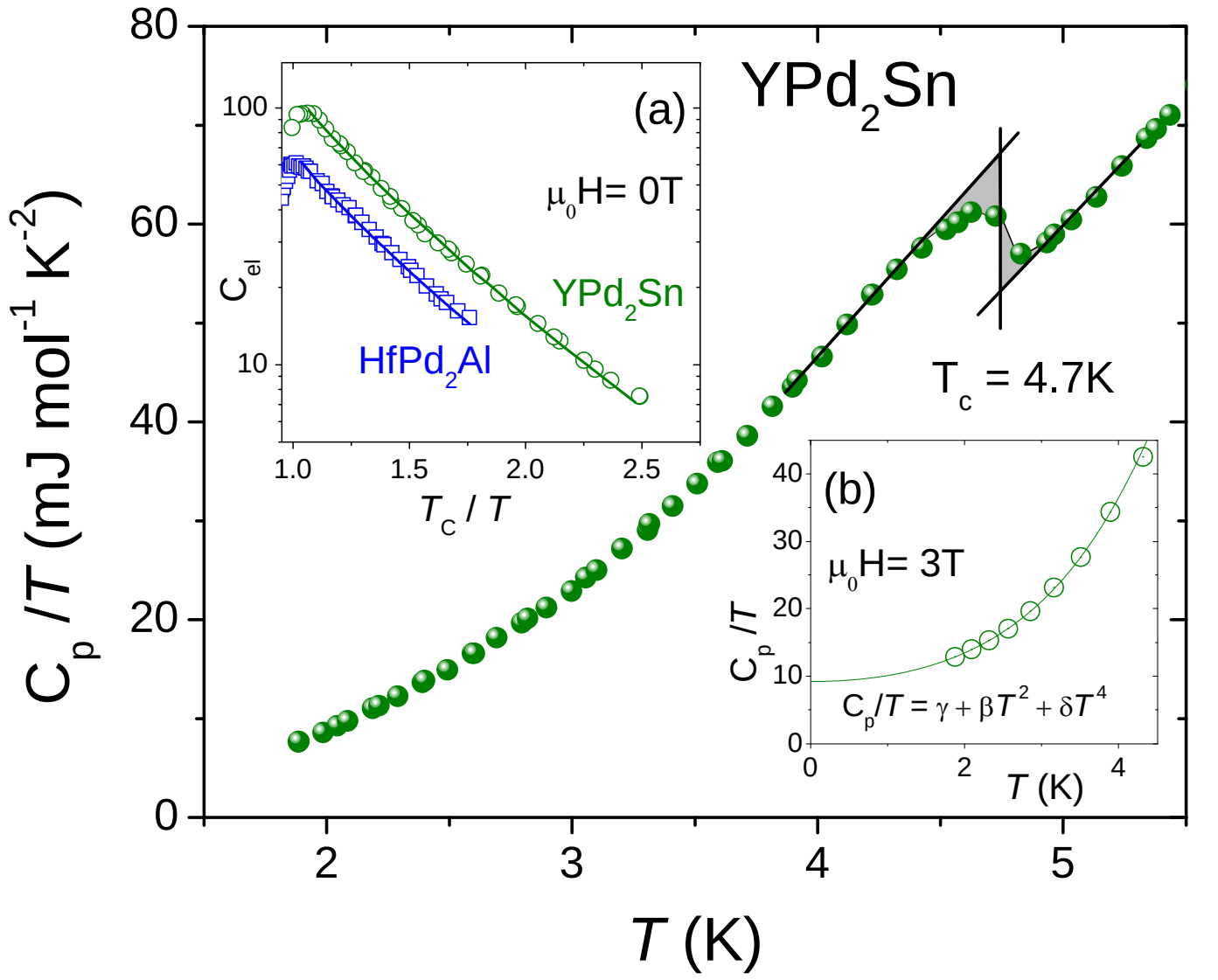

Figure 6.

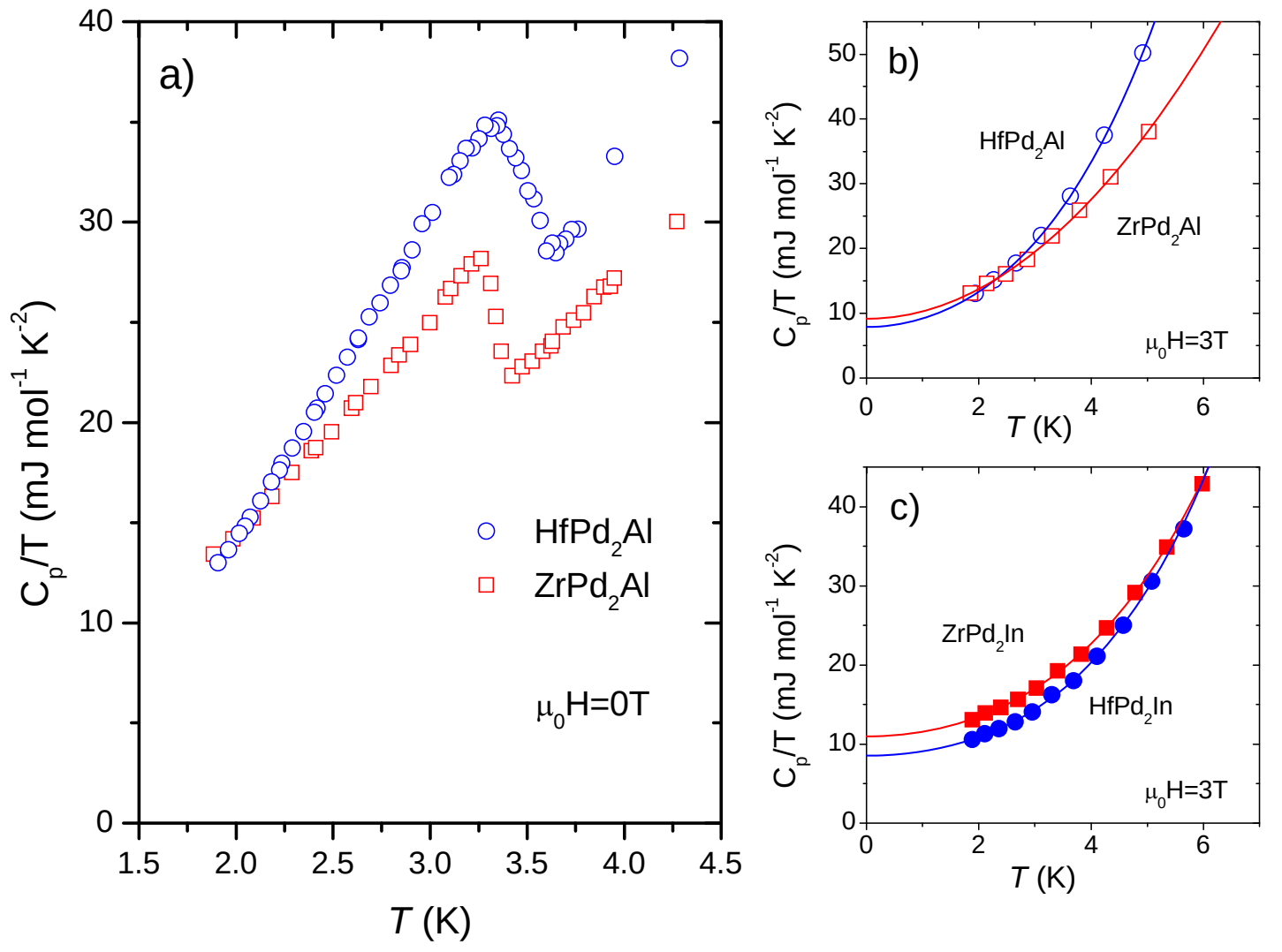

Figure 7.

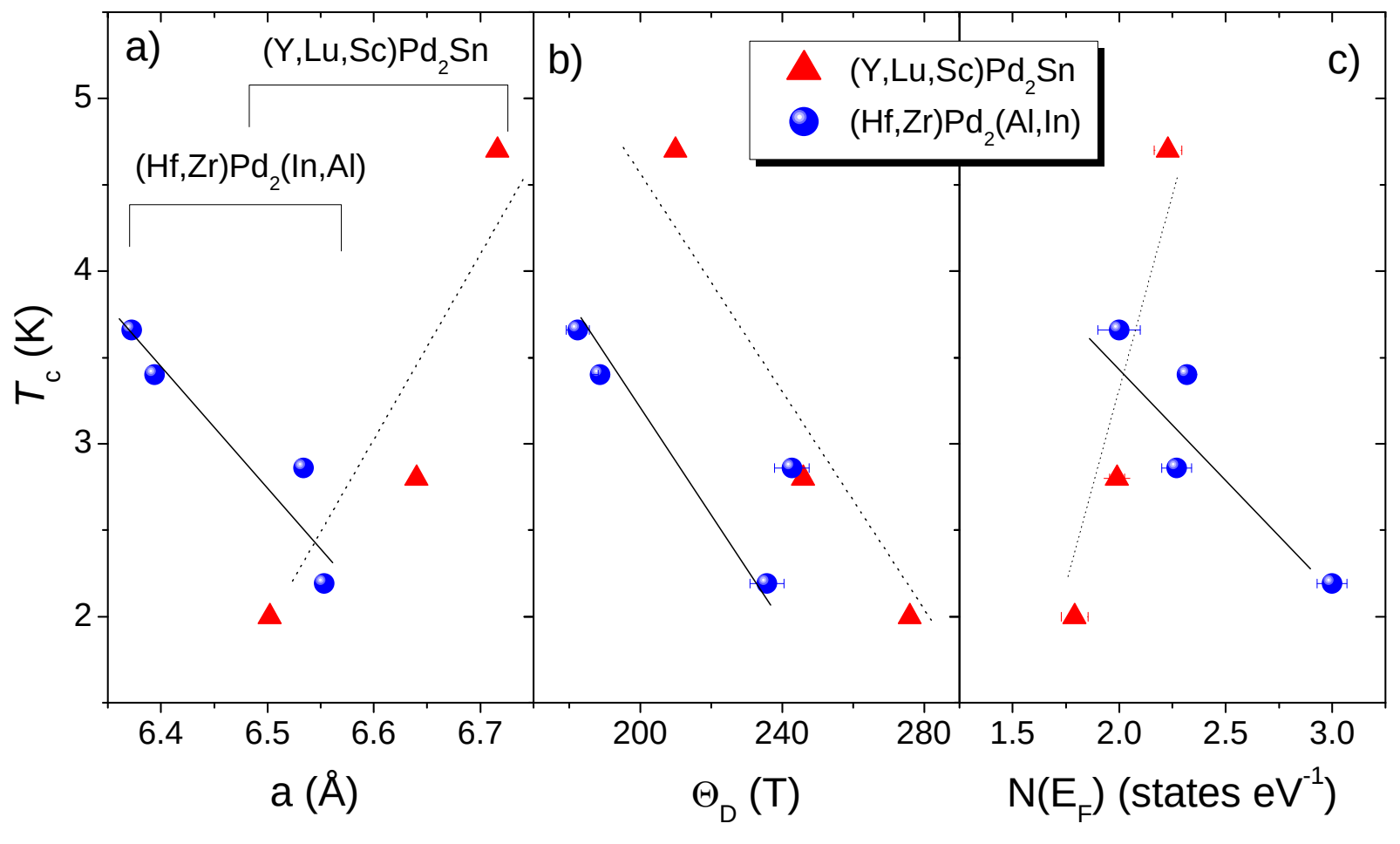

Figure 8.

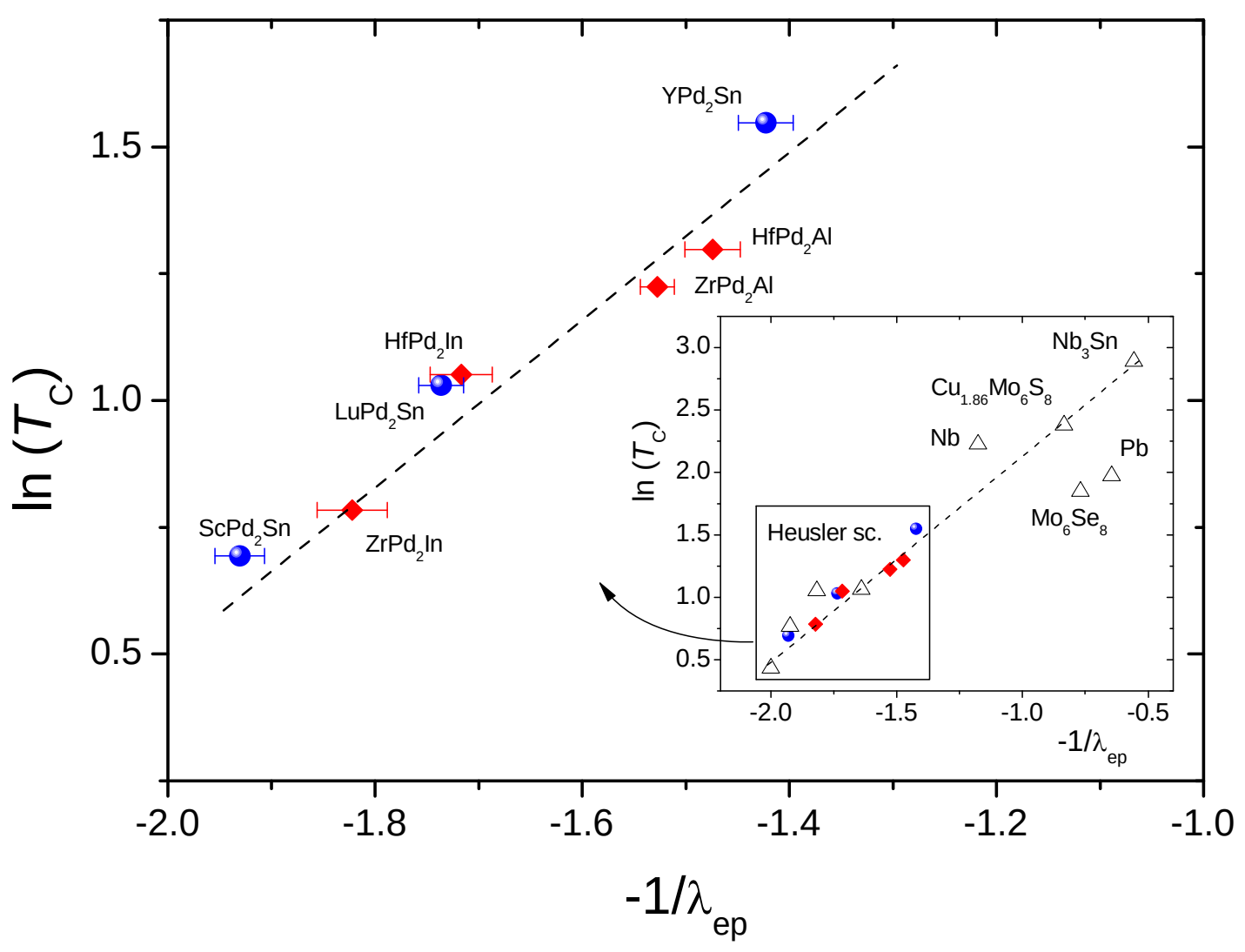

Figure 9.

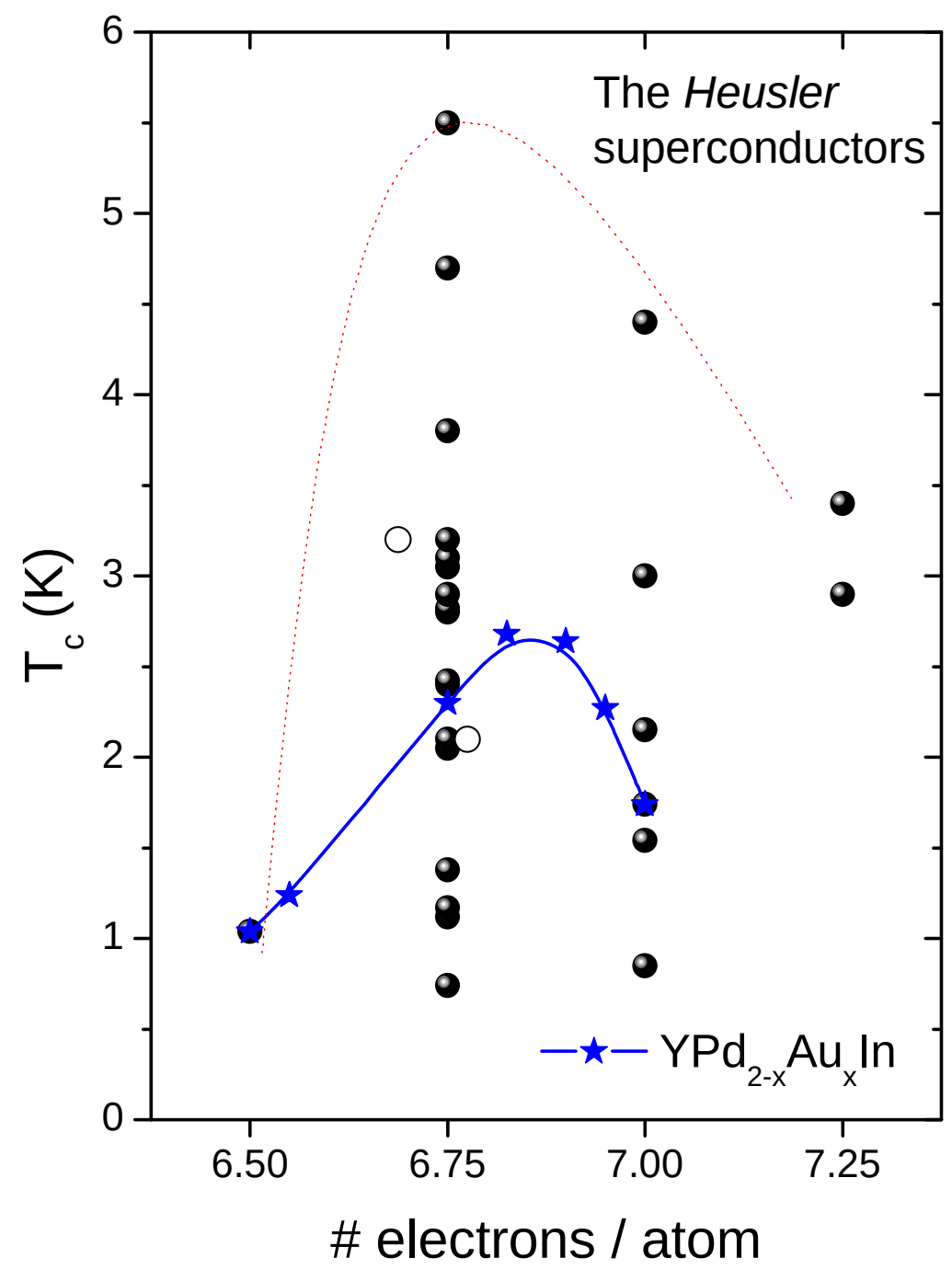

Figure 10.

[19] S. Waki, Y. Yamaguchi, and K. Mitsugi, Journal of the Physical Society of Japan **54**, 1673 (1985).

[20] H. A. Kierstead, B. D. Dunlap, S. K. Malik, A. M. Umarji, and A. G. K. Shenoy, Phys. Rev. B **32**, 135 (1985).

[21] B. T. Matthias, Physical Review **92**, 874 (1953).

[22] B. T. Matthias, Physical Review **97**, 74 (1955).

[23] J. Rodriguez-Carvajal, Physica B: Physics of Condensed Matter **192**, 55 (1993).

[24] T. Klimczuk, T. M. McQueen, A. J. Williams, Q. Huang, F. Ronning, E. D. Bauer, J. D. Thompson, M. A. Green, and R. J. Cava, Phys. Rev. B **79**, 012505 (2009).

[25] J. Winterlik, G. H. Fecher, C. Felser, M. Jourdan, K. Grube, F. Hardy, H. Von Löhneysen, K. L. Holman, and R. J. Cava, Phys. Rev. B 78, 184506 (2008).

[26] We use the temperature at which the resistivity reaches half its normal-state value, $\rho(T)=1/2\ \rho_{res}$, to define the midpoint $T_c$.

[27] N. R. Werthamer, E. Helfand, and P. C. Hohenberg, Physical Review **147**, 295 (1966).

[28] Dirty limit was deduced from the fact that the ratio $\xi_0/l$ is greater than 1, where l is quasiparticle mean free path and was calculated from useful equation $l = 2.372\times10^{-14}\,\frac{(m^*/m_e)^2 V_M^2}{N(E_F)^2\rho}$ derived by Y. Singh, et al., Phys. Rev. B **82**, 144532 (2010). Taking values for $YPd_2Sn$: $V_M$ = 45.6 cm$^3$ mol$^{-1}$, $N(E_F)$ = 2.23 states eV$^{-1}$, $\rho$ = 18 μΩ cm, and assuming $m^*/m_e = 1$, we obtained l = 550 nm, and $\xi_0/l = 3.45$, which is the lowest value for all studied compounds.

[29] S. K. Malik, A. M. Umarji, and G. K. Shenoy, Phys. Rev. B 34, 3144 (1986).

[30] B. Mühlschlegel, Z. Phys. 155, 313 (1959)

[31] $\mu^*$ = 0.13 falls in between the used in the literature range 0.1 and 0.15. The same value was also used by A. B. Karki, Y. M. Xiong, N. Haldolaarachchige, S. Stadler, I. Vekhter, P. W. Adams, D. P. Young, W. A. Phelan, and Julia Y. Chan, Phys. Rev. B 83, 144525 (2011).

[32] W. L. McMillan, Physical Review **167**, 331 (1968).

[33] J. L. Jorda, M. Ishikawa, and J. Muller, Journal of the Less-Common Metals **107**, 321 (1985).

[34] S. K. Malik, A. M. Umarji, and G. K. Shenoy, Phys. Rev. B **32**, 4426 (1985).

[35] observed for $Y_{0.96}Pd_{2.08}Sn_{0.96}$ as reported in ref. 33.

[36] C. K. Poole, H. A. Farach, and R. J. Creswick, Handbook of Superconductivity (Academic Press, 2000), ISBN 0125614608.

[37] J. Winterlik, G. H. Fecher, A. Thomas, and C. Felser, Phys. Rev. B **79**, 064508 (2009).